\documentclass{aastex62}

\usepackage{xspace}

\newcommand{\HST}{{\it HST}\xspace}
\newcommand{\JWST}{{\it JWST}\xspace}
\newcommand{\WFIRST}{{\it WFIRST}\xspace}

\newcommand{\foutl}{\ensuremath{f^{\mathrm{outl}}}\xspace}
\newcommand{\foutlij}{\ensuremath{f^{\mathrm{outl}}_{ij}}\xspace}
\newcommand{\muinlij}{\ensuremath{\mu^{\mathrm{inl}}_{ij}}\xspace}
\newcommand{\muoutlij}{\ensuremath{\mu^{\mathrm{outl}}_{ij}}\xspace}

\newcommand{\paramunits}{$\frac{\mathrm{e}^-}{\mathrm{pix} \cdot \mathrm{s}}$\xspace}

\submitjournal{RN}

\shorttitle{Erratic WFC3 IR Pixels}
\shortauthors{Currie and Rubin}

\begin{document}

\title{Characterization of Unstable Pixels Using a Mixture Model: Application to \HST WFC3 IR}

\author[0000-0003-3429-4142]{Miles Currie}
\affil{Space Telescope Science Institute \\
3700 San Martin Dr.
Baltimore, MD 21218, USA}
\email{mcurrie@stsci.edu}

\author[0000-0001-5402-4647]{David Rubin}
\affil{Space Telescope Science Institute \\
3700 San Martin Dr.
Baltimore, MD 21218, USA}
\affil{Lawrence Berkeley National Laboratory\\
1 Cyclotron Road
Berkeley, CA 94720, USA}
\maketitle

Many IR datasets are taken with two dithers per filter, complicating the automated recognition of pixels with unstable response. Much data from the \HST cameras NICMOS and WFC3 IR fall into this category, and future \JWST and \WFIRST data are likely to as well. It is thus important to have an updated list of unstable pixels built from many datasets. We demonstrate a simple Bayesian method that directly estimates the fraction of the time the output of each pixel is unstable. The last major update for WFC3 IR was a 2012 instrument science report \citep[ISR~WFC3~2012-10,][]{hilbert12}, so we compute a new list. Rather than reproduce the old analysis on newer data, we use our new method. By visual inspection, our method identifies unstable pixels with better purity and completeness.

We create our maps of unstable pixels using the WFC3 IR dark frames. Dark frames are ideal for this purpose because they have simple structure and low count rates for high sensitivity. In this note, we focus on two epochs: 2017 to make an updated unstable pixel list and 2009-2012 to compare against \citet{hilbert12}. For the \citet{hilbert12} comparison, we take the same 117 files used in that analysis and apply our method. For 2017, we exclude images taken within one hour of the South Atlantic Anomaly passage, leaving 177 darks to create our updated list of unstable pixels.


Our key quantity is the fraction of the time a pixel is unstable. We model the flux value in each pixel's time-series as the following Gaussian mixture with likelihood given by
\begin{equation} \label{eq:likelihood}
L_{ij} = \prod_{k=1}^{N_\mathrm{images}} \left[ (1 - \foutlij) \; \mathcal{N}(\muinlij, \sigma_{ijk}^{2} + 0.02^2) + \foutlij \; \mathcal{N}(\muoutlij, (\sigma^{\mathrm{outl}}_{ij})^2) \right] \;.
\end{equation}
Each pixel ($i, j$) is modeled independently using the following parameters: 1) \foutlij is our estimate of the fraction of the time a pixel ($i, j$) behaves erratically. 2) \muinlij is our estimate of the dark current for each pixel (assumed to be constant with time). 3) $\sigma_{ijk}$ are the \texttt{CALWF3} pipeline uncertainties for each pixel (in image $k$). 4) Broadband science data has larger uncertainties than the darks, so erratic behavior with an amplitude smaller than 0.02 \paramunits is unlikely to impact science data. We thus include an uncertainty floor added in quadrature with the \texttt{CALWF3} uncertainties, which we fix in size to 0.02 \paramunits. 5) \muoutlij describes the mean of the distribution of any erratic behavior. 6) $\sigma^{\mathrm{outl}}_{ij}$ describes the width of any erratic behavior.

We sampled from the model in Stan \citep{carpenter17} using PyStan (\url{https://pystan.readthedocs.io}), running four chains and ensuring that $\hat{R}$ diagnostic \citep{gelman92} was within 0.05 of 1.0. In the unusual case it was not (indicating poor sampling), we reran with new randomly selected initial conditions.

Other than the small uncertainty floor in the inlier distribution, Equation~\ref{eq:likelihood} is symmetric in inliers and outliers. To identify the outlier pixel values with the outlier distribution and remove this symmetry, we require $\sigma_{\mathrm{outl}} \geq 0.05$. We assume the following weak priors (all in units of \paramunits):

\begin{eqnarray}
\mu^{\mathrm{inl}} &\sim & \mathcal{N}(0, 1^2)  \\ 
\sigma^{\mathrm{outl}} &\sim & \mathcal{N}(0, 1000^2)  \\ 
\mu^{\mathrm{outl}} &\sim & \mathcal{N}(0, 1000^2)  \\ 
\foutl &\sim& \mathcal{U} (0, 1) 
\end{eqnarray}
In principle, one could assume a prior concentrated around zero for $\foutl$, as the vast majority of pixels do not show (detectable) erratic behavior. But the $\foutl$ are well enough measured (typical uncertainties are $\sim$ 0.02 - 0.1) that this prior would only have a modest effect on the posteriors.

\begin{figure*}[h]
\centering
\includegraphics[width=0.8\textwidth]{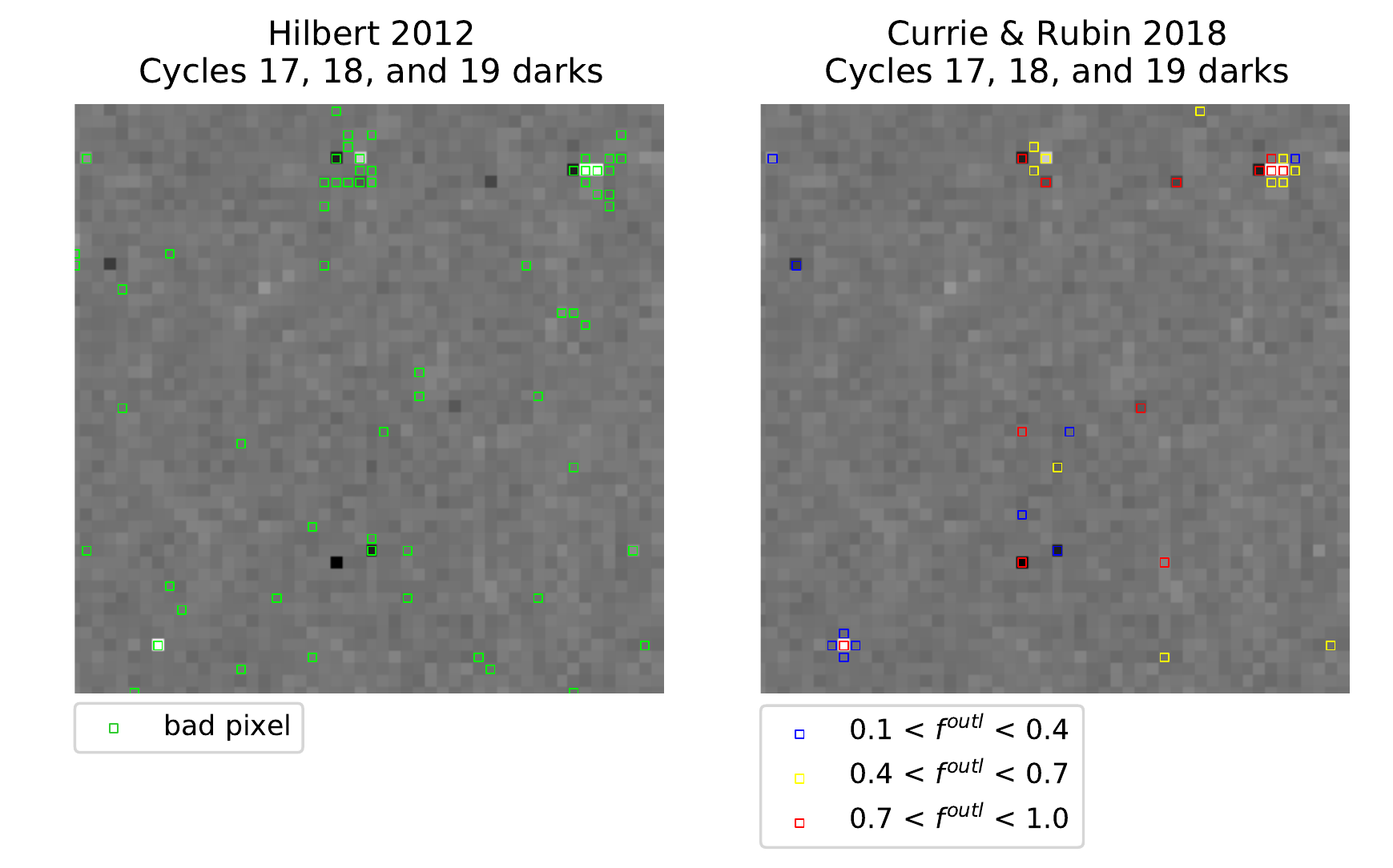}
\caption{Comparison of \citet{hilbert12} (left) and our (right) flagged unstable pixels on the same randomly selected file (idgc01i5q\_flt.fits), which was not used to build the model. Both models were run on the same data from \HST Cycles 17, 18, and 19. Our model flags roughly half the amount of pixels that Hilbert flags. Our unstable pixels are color coded by \foutl values; the pixels closer to one are more frequently unstable. Our model is visually better at identifying unstable pixels.}
\label{fig:onlyfigure}
\end{figure*}

We present our recovered results and a comparison to the results of \citet{hilbert12} in Figure~\ref{fig:onlyfigure}. The left panel shows a cutout from a randomly selected WFC3 IR image from 2017 (idgc01i5q\_flt.fits), with \citet{hilbert12}'s unstable pixels outlined in green. The right panel shows the same cutout with our flagged unstable pixels outlined in colors corresponding to the pixel's outlier fraction.  Hilbert's algorithm gives 21,442 unstable pixels ($\sim$ 2\% of all WFC3 pixels) while our model gives 10,551 unstable pixels ($\sim$ 1\% of all WFC3 pixels) over a range of \foutl values greater than 0.1. Our algorithms agree on 5,681 pixels. For the 2017 darks, our algorithm flags 14,935 ($\sim$ 1.5\%) unstable pixels and agrees with our 2012 data analysis on 7,520 pixels. It agrees with \citet{hilbert12} on 6,211 pixels. Our model and both the 2012 and 2017 pixel lists can be found at \url{https://zenodo.org/badge/latestdoi/139165857}. Finally, we note that Ben Sunnquist at STScI is computing a different update to the \citet{hilbert12} model, and it shows much better overlap with ours (Ben Sunnquist, private communication). 

\acknowledgements
This work was supported by HST-GO 14808 and 15363, and a NASA \WFIRST Science Investigation Team. We thank Susana Deustua and Ben Sunnquist for their feedback. 

\bibliographystyle{apj}
\bibliography{rnaas}



\end{document}